\documentclass[pra, aps, twocolumn, groupedaddress, showpacs, superscriptaddress]{revtex4-2}
\usepackage{amssymb,amsmath,amsthm,color,graphicx,times,graphicx}
\usepackage{hyperref}
\usepackage{subfigure}
\usepackage{times}
\usepackage{bbold}
\usepackage{color}

\theoremstyle{plain}

\theoremstyle{definition}

\begin{document}
\title{Enhancing the estimation precision of an unknown phase shift in multipartite Glauber coherent states via skew information correlations and local quantum Fisher information}
\author{Mehdi El Bakraoui}\email{mehdielbekraoui@gmail.com }\affiliation{LPHE-Modeling and Simulation, Faculty of Sciences, Mohammed V University in Rabat, Rabat, Morocco.}
\author{Abdallah Slaoui}\email{abdallah.slaoui@um5s.net.ma}\affiliation{LPHE-Modeling and Simulation, Faculty of Sciences, Mohammed V University in Rabat, Rabat, Morocco.}\affiliation{Centre of Physics and Mathematics, CPM, CNESTEN, Rabat, Morocco.}
\author{Hanane El Hadfi}\email{hanane.elhadfi@gmail.com}\affiliation{LPHE-Modeling and Simulation, Faculty of Sciences, Mohammed V University in Rabat, Rabat, Morocco.}
\author{Mohammed Daoud}\email{m\_daoud@hotmail.com}\affiliation{Department of Physics, Faculty of Sciences, University Ibn Tofail, Kenitra, Morocco.}\affiliation{Abdus Salam International Centre for Theoretical Physics (ICTP), Strada Costiera, 11 I-34151, Trieste, Italy.}

\begin{abstract}
Local quantum uncertainty (LQU) and local quantum Fisher information (LQFI) are both two tools used to capture purely quantum correlations in multi-partite quantum systems. In this paper, we study these quantifiers in the case of multipartite Glauber coherent state which include the GHZ (Greenberger-Horne-Zeilinger) and Werner states. We perform a comparative study between LQFI and LQU in an isolated system. Besides, by using the Kraus operator representation, we study the behavior of these quantifiers on the dephasing channel to investigate their performances under the decoherence effect. In addition, the robustness to the decoherence effect of these two quantifiers is studied. We further examine the situation involving the multipartite Glauber coherent state to decide the sensitivity of the probe state as a resource for quantum estimation protocols.
\par
\vspace{0.25cm}
\end{abstract}
\date{\today}

\maketitle
\section{Introduction}

Quantum information is a relevant field of information and communication science and technology that is undergoing rapid development. The presence of non-local correlations is one of the most prominent manifestations that distinguishes the quantum world from its classical counterpart, and it is a kind of important physical resource for quantum information processing \cite{Watrous2018,Renes2014}. Over the past few decades, many important advances have been achieved in this field with many powerful applications exploiting quantum properties such as quantum superposition and entanglement, to perform quantum tasks that are not possible using the laws of classical physics. Quantum entanglement has been at the center of fundamental discussions since the work of Schrödinger \cite{Schrodinger1935} and the paper by Einstein, Podolsky and Rosen \cite{Einstein1935}, who proposed it as a way to quantify the quantum correlations present in diverse types of quantum systems. However, it is now well known that there are more general quantum correlations beyond entanglement, which manifest themselves even in some separable mixed states \cite{Henderson2001,Ollivier2001}. In this respect, the characterization and quantification of quantum correlations beyond entanglement have attracted much attention and have become one of the most extensively studied problems in recent years \cite{Luo2008,SlaouiS2020,Slaoui20182,ShaukatS2020}.\par

In general, quantum correlations present an important resource in several quantum information processing protocols and they have been shown useful in the field of quantum metrology. In fact, nonclassical correlations have now been shown to be a useful feature for extending the range of nonclassical estimation protocols \cite{Nielsen2002,Paris2009}. Fragile quantum features such as quantum correlations are used to improve the accuracy of parameter estimation, and also, the gain of information that becomes resistant to noise is essential to promise these quantum features in quantum metrology \cite{Sen2016,Petz2011}. To estimate an unknown parameter in a quantum system, one usually prepares a probe that interacts with the system, and then measures the probe. If the physical mechanism controlling the dynamics of the system is known, we can approximate the value of the parameter by comparing the input and output states of the probe. Since quantum states can rarely be distinguished with certainty, there is usually an existing statistical uncertainty in this approximation. To reduce this uncertainty, we can use $N$ identical and independent probes, measure them and average the results. According to the central limit theorem \cite{Paris2009}, the error on the average decreases as $\Delta / \sqrt{N}$ for large values of $N$, with $\Delta^{2}$ is the variance of the measurement results associated with each probe. Using the same physical resources with the addition of quantum effects, such as entanglement or squeezing, an even better precision can often be achieved with a customary  $\sqrt{N}$ enhancement \cite{Giovannetti2011,Toth2014,Szczykulska2016}.\par

Here, we  investigate the effects of quantum correlations existing in even and odd suppositions of Glauber coherent states to ameliorate the precision of an unknown parameter \cite{Everett1957}. The accuracy of the quantum measurement is related to Heisenberg scaling. In order to reach this accuracy, we should use the entanglement state as a source of probes. The realization of these benefits therefore depends on the preparation of special non-classical states that encode the probe state setting in such a way that it can be determined with a precision that exceeds the standard quantum limit \cite{Helstrom1969}. In a quantum state, the ultimate limit on the obtainable accuracy is provided by the quantum Cramér-Rao bound via the quantum Fisher information. A parameter $\theta$ can be estimated through an (unbiased) estimator $\hat{\theta}$ and the limit of the precision of their measurement is generally fixed by the Cramér-Rao quantum limit which reads
\begin{equation}
\operatorname{Var}(\hat{\theta}) \geq \frac{1}{n \mathcal{F}(\rho, H)},
\end{equation}
where $n$ denotes the number of independent repetitions. It is clear that the inverse of the QFI depicts the lower error limit in statistical estimation of an unknown parameter \cite{SlaouiB2019}.\par

In quantum estimation theory (QET), the use of quantum Fisher information as a quantifier provides a quantitative ingredient to determine the precision of an unknown parameter. Its inverse gives the Cramer-Rao bound \cite{Paris2009,Helstrom1969}. It bounds the achievable precision in parameter estimation with a quantum system. Early theoretical efforts in quantum metrology centered around designing quantum states that saturate this bound. In fact, to understand the role of quantum correlation beyond entanglement in a black box quantum metrology task, two quantum correlation quantifiers have recently been introduced. The first quantifier is the local quantum uncertainty (LQU) \cite{Girolami2013} while the second is the local quantum Fisher information (LQFI) \cite{Kim2018}. These measures are discord-type quantifiers and they are easily computable in comparison with other quantum correlations like entropic quantum discord which involves complicated optimization procedures that are not easy to solve in general \cite{Huang2014}. Very importantly, LQU and LQFI of a mixed bipartite probe state ensure the minimum accuracy quantified by quantum Fisher information in the optimal phase estimation protocol. Moreover, closed formulas exist only for $2\otimes d$ bipartite quantum states \cite{SlaouiB2019}. Interestingly, a comparative study between geometric quantum discord and LQU was performed in \cite{Cordero2021} and it was found that these two discord-like measures of quantum correlation are equivalent for ($2\otimes d$)-dimensional bipartite quantum systems. Moreover, the geometric quantum discord has shown tremendous importance in the phase estimation framework when we use NOON states as probe sources in lossy non-symmetric environments. In fact, various geometric measures of quantum discord have been suggested in the literature \cite{Luo2010,Paula2013}. To prove its intrinsic relation with LQU, we need to use the alternative measure with more informational significance by employing the square root of a density operator, as proposed in Ref.\cite{Chang2013}.\par

This work is organized as follows. In Sec.\ref{I}, we introduce the realization of logical qubits via multipartite coherent states  and  the density matrix for the system we need to study. In Sec.\ref{II},  the local quantum Fisher information is derived for multipartite coherent states. We give the explicit expression of local quantum Fisher information. In Sec.\ref{III}, the expression of  local quantum uncertainty for multipartite coherent states is given. Sec.\ref{V} is devoted to the analysis of the dynamical evolution of the quantum Fisher information and local quantum uncertainty under a dephasing channel to examine the decoherence effects on the estimation protocols. Concluding remarks close this paper.
\section{Realization of logical qubits via multipartite coherent state superpositions}\label{I}
Coherent states are important in many fields of physics \cite{Monroe1996,Zhang1990}. This became widely recognized during the 1960’s thanks to the works of Glauber \cite{Glauber1963}, Klauder \cite{Klauder1985} and Perelomov \cite{Perelomov1972,Gazeau2009,Gazeau2000}. For the harmonic oscillator, the coherent states are given by the unitary displacement of on the ground state vector $\left|0\right\rangle$
\begin{equation}
D(\alpha)|0\rangle=\exp \left[-\frac{1}{2}|\alpha|^{2}\right] \sum_{n} \frac{\alpha^{n}}{\sqrt{n !}}|n\rangle \equiv|\alpha\rangle, 
\end{equation}
with the displacement operator given by
\begin{align}
D(\alpha)&=\exp\left[\alpha a^{\dagger}-\alpha^{*} a\right]\notag\\&=\exp \left[-|\alpha|^{2} / 2\right] \exp \left[\alpha a^{\dagger}\right] \exp \left[-\alpha^{*} a\right],
\end{align}
where $\alpha$ is a complex dimensionless amplitude of the coherent state and $|n\rangle$ is a Fock state. It is interesting to note that
the generalization of Glauber method has been extended to arbitrary Lie groups by Perelomov (see Refs. \cite{Perelomov1986} and later reviews \cite{Sanders2012}). One simply applies the generalized displacement operator, which is the unitary exponentiation of the factor algebra, on to an extremal state. To simplify our purpose, we shall essentially focus on the superposition of multipartite coherent states  $|\alpha\rangle$ and $|-\alpha\rangle$ of the form
\begin{equation}
|\psi\rangle \simeq N\left(|\alpha,\alpha,......,\alpha\rangle+e^{i m\pi}|-\alpha,........,-\alpha\rangle\right),\label{3}
\end{equation}
where $m=0$ for symmetric state and $m=1$ for antisymmetric state.
The normalization factor $N$ reads
\begin{equation}
N=\left[2+2 p^{n} \cos m \pi\right]^{-1 / 2}
\end{equation}
and $p=\langle\alpha \mid-\alpha\rangle = \exp \left[-2|\alpha|^{2}\right]$ is the overlap 
between two single states. We shall essentially study the asymptotic bounds, the first is when $p \rightarrow 0$ and the second case is  $p \rightarrow 1$. When $p \rightarrow 0$,  it's obvious that the states  $|\alpha\rangle$ and $|-\alpha\rangle$ are orthogonal
and $N = \frac{1}{\sqrt{2}}$. So, it's acting like GHZ type state as 
\begin{equation}
\begin{aligned}
|\mathrm{GHZ}\rangle_{n}=& \frac{1}{\sqrt{2}}(|\alpha\rangle \otimes|\alpha\rangle \otimes \cdots \otimes|\alpha\rangle\\
&\left.+e^{i m \pi}|-\alpha\rangle \otimes|-\alpha\rangle \otimes \cdots \otimes|-\alpha\rangle\right).
\end{aligned}
\end{equation}
In the second limiting case, when $p \rightarrow 1$ (or $\alpha\rightarrow0$), we should 
specify separately the case when $m=0$ and  $m=1$. For $m=0$, our multipartite superposition state $|\psi\rangle$ approaches the ground state;
\begin{equation}
|0,\alpha, n\rangle \sim|\alpha\rangle \otimes|\alpha\rangle \otimes \cdots \otimes|\alpha\rangle.
\end{equation}
It is interesting to note that when $p \rightarrow 1$, the state (\ref{3}) with $m$ odd reduces to a superposition similar to the so-called Werner state, i.e., 
\begin{equation}
\begin{aligned}
|\mathrm{W}\rangle_{n}=& \frac{1}{\sqrt{n}}(|-\alpha\rangle \otimes|\alpha\rangle \otimes \cdots \otimes|\alpha\rangle\\
&+|\alpha\rangle \otimes|-\alpha\rangle \otimes \ldots \otimes|\alpha\rangle \\
&+\cdots+|\alpha\rangle \otimes|\alpha\rangle \otimes \cdots \otimes|-\alpha\rangle).
\end{aligned}
\end{equation}
To study the LQFI and LQU in the Glauber coherent state(\ref{3}), we split the system into two differents ways. The first gives pure bipartite states and the second gives mixed states. In the first bipartite splitting, we write (\ref{3}) as a bipartite quantum state
\begin{equation}
|\psi\rangle=N\left(|\alpha\rangle_{k} \otimes|\alpha\rangle_{n-k}+e^{i m \pi}|-\alpha\rangle_{k} \otimes|-\alpha\rangle_{n-k}\right),\label{8}
\end{equation}
where $k=1,2,......,n-1$. To express the state (\ref{8}) as logical qubit, we introduce the orthogonal basis $\left\{|0\rangle_{k},|1\rangle_{k}\right\}$ defined as:
\begin{equation}
|0\rangle_{k}=\frac{|\alpha\rangle_{k}+|-\alpha\rangle_{k}}{\sqrt{2\left(1+p^{k}\right)}}, \quad|1\rangle_{k}=\frac{|\alpha\rangle_{k}-|-\alpha\rangle_{k}}{\sqrt{2\left(1-p^{k}\right)}},\label{9}
\end{equation}
for the qubit $1$ (the first subsystem). Similarly, for the second subsystem, the orthogonal basis $\left\{|0\rangle_{n-k},|1\rangle_{n-k}\right\}$ given by
\begin{equation}
\begin{aligned}
&|1\rangle_{n-k}=\frac{|\alpha\rangle_{n-k}-|-\alpha\rangle_{n-k}}{\sqrt{2\left(1-p^{n-k}\right)}},\\
&|0\rangle_{n-k}=\frac{|\alpha\rangle_{n-k}+|-\alpha\rangle_{n-k}}{\sqrt{2\left(1+p^{n-k}\right)}}.\label{10}
\end{aligned}
\end{equation}
Substituting the equations (\ref{9}) and (\ref{10}) into equation (\ref{8}), the state (\ref{8}) becomes
\begin{equation}
|\psi\rangle=\sum_{i=0,1} \sum_{j=0,1} C_{i, j}|i\rangle_{k} \otimes|j\rangle_{n-k},
\end{equation}
with 
\begin{equation}
\begin{array}{ll}
C_{0,0}=N\left(1+e^{i m \pi}\right) a_{k} a_{n-k}, \\ C_{0,1}=N\left(1-e^{i m \pi}\right) a_{k} b_{n-k}, \\
C_{1,0}=N\left(1-e^{i m \pi}\right) a_{n-k} b_{k}, \\ C_{1,1}=N\left(1+e^{i m \pi}\right) b_{k} b_{n-k},\label{12}
\end{array}
\end{equation}
in the basis $$\left\{|0\rangle_{k} \otimes|0\rangle_{n-k,}|0\rangle_{k}\otimes |1\rangle_{n-k},|1\rangle_{k} \otimes|0\rangle_{n-k},|1\rangle_{k} \otimes|1\rangle_{n-k}\right\}.$$
The coefficients occurring in (\ref{12}) are given by
$$
a_{l}=\sqrt{\frac{1+p^{l}}{2}}, \quad b_{l}=\sqrt{\frac{1-p^{l}}{2}} \quad \text { for } l=k, n-k .
$$
The density matrix associated with the state (\ref{3}) writes
\begin{equation}
\begin{aligned}
\rho=&|\psi\rangle\langle\psi|= N^{2}(|\alpha, \alpha\rangle\langle\alpha, \alpha|+|-\alpha,-\alpha\rangle\langle-\alpha,-\alpha|\\
&\left.+e^{i m \pi} q|-\alpha,-\alpha\rangle\left\langle\alpha, \alpha\left|+e^{-i m \pi} q\right| \alpha, \alpha\right\rangle\langle-\alpha,-\alpha|\right),
\end{aligned}
\end{equation}
with $q\equiv p^{n-2}$. Hence, the reduced density matrix for the subsystems containing the the qubits $1$ and $2$ writes as 
\begin{equation}
\begin{aligned}
\rho_{12}=& \operatorname{Tr}_{3,4, \ldots, n}(|\alpha, m, n\rangle\langle\alpha, m, n|) \\
=& N^{2}(|\alpha, \alpha\rangle\langle\alpha, \alpha|+|-\alpha,-\alpha\rangle\langle-\alpha,-\alpha|\\
&\left.+e^{i m \pi} q|-\alpha,-\alpha\rangle\left\langle\alpha, \alpha\left|+e^{-i m \pi} q\right| \alpha, \alpha\right\rangle\langle-\alpha,-\alpha|\right).\label{14}
\end{aligned}
\end{equation}
To convert the density matrix $\rho_{12}$ into a two-qubit system, we consider an orthogonal pair of states $\{|\mathbf{0}\rangle,|\mathbf{1}\rangle\}$ defined by
\begin{equation}
|\alpha\rangle \equiv a|0\rangle+b|1\rangle, \hspace{1cm}\left|-\alpha\right\rangle \equiv a\left| 0\right\rangle-b\left|1\right\rangle,
\end{equation}
where
\begin{equation}
a=\sqrt{\frac{1+p}{2}}, \hspace{1cm} b=\sqrt{\frac{1-p}{2}}.
\end{equation}
The logical qubits $|\mathbf{0}\rangle$ and $|\mathbf{1}\rangle$ correspond to the even and odd coherent states given respectively by
\begin{equation*}
|0\rangle=\frac{1}{\sqrt{2+2 p}}(|\alpha\rangle+|-\alpha\rangle), \quad|1\rangle=\frac{1}{\sqrt{2-2 p}}(|\alpha\rangle-|-\alpha\rangle).
\end{equation*}
The second bipartite scheme we shall consider in this work corresponds to bipartite states extracted from the multipartite state (\ref{3}). In this way, we consider two quantum states obtained from (\ref{3}) by tracing out the other $(n-2)$ modes. Then the density matrix (\ref{14}) takes, in the computational basis $\{|\mathbf{0 0}\rangle,|\mathbf{0 1}\rangle,|\mathbf{1 0}\rangle,|\mathbf{1 1}\rangle\}$, the form 
\begin{equation}
\rho_{12}=N^{2}\left(\begin{array}{cccc}
\eta_{a} & 0 & 0 & \eta_{+} \\
0 & \eta_{-} & \eta_{-} & 0 \\
0 & \eta_{-} & \eta_{-} & 0 \\
\eta_{+} & 0 & 0 & \eta_{b}
\end{array}\right),\label{Matrix12}
\end{equation}
with
\begin{equation*}
\eta_{a}=2 a^{4}(1+q \cos m \pi),\hspace{0.5cm}\eta_{b}=2 b^{4}(1+q \cos m \pi),
\end{equation*}
and
\begin{equation*}
\eta_{\pm}=2 a^{2} b^{2}(1\pm q\cos m \pi).
\end{equation*}
\section{Local Quantum fisher information for bipartite coherent states}\label{II}
In quantum estimation theory, the quantum Fisher information
(QFI) is recognized as an important tool to determine the ultimate accuracy in parameter estimation scenarios. The key concept in quantum metrology is the quantum Cramér Rao bound in estimating a given unknown physical parameter \cite{Paris2009,Szczykulska2016}. In a generic estimation protocol, an initial quantum state $\hat{\rho}(0)$ that undergoes unitary evolution $\hat{U}_{\theta}=\exp (-i \hat{H} \theta)$, the issue is how precisely can the  parameter $\theta$ be estimated? The answer is given by the quantum Cramér-Rao bound (QCRB) which places a lower bound on the sensitivity \cite{BakmouS2019,Laghmach2019}, i.e. $\Delta \theta \geq 1 / \sqrt{\mathcal{F}}$ in term of quantum Fisher information given by
\begin{equation}
\mathcal{F}\left(\rho_{\theta}\right)=\frac{1}{4} \operatorname{Tr}\left[\rho_{\theta} L_{\theta}^{2}\right],
\end{equation}
 where the symmetric logarithmic derivative $L_{\theta}$ is defined as the solution of the equation
$\frac{\partial \rho_{\theta}}{\partial \theta}=\frac{1}{2}\left(L_{\theta} \rho_{\theta}+\rho_{\theta} L_{\theta}\right) .$
The parametric states $\rho_{\theta}$ can be obtained from an initial probe state $\rho$ subjected to $\theta$-dependent unitary transformation $U_{\theta}=e^{i H \theta}$ and generated by a Hermitian operator $H$, i.e., $\rho_{\theta}=U_{\theta}^{\dagger} \rho U_{\theta}$. In this case, the quantum Fisher information  $\mathcal{F}\left(\rho_{\theta}\right) $ \cite{Paris2009}, that we denote by $\mathcal{F}(\rho, H)$, is given by
\begin{equation}
\mathcal{F}(\rho, H)=2 \sum_{i, j} \frac{\left(\lambda_{i}-\lambda_{j}\right)^{2}}{\lambda_{i}+\lambda_{j}}\left|\left\langle \psi_{i}|\hat{H}| \psi_{j}\right\rangle\right|^{2}\label{F},
\end{equation}
where $\lambda_{i} $ and $ \left|\psi_{i}\right\rangle$ are the eigenvalues and eigenvectors of $\rho$, with $\rho=\sum_{i=1} \lambda_{i}\left|\psi_{i}\right\rangle\left\langle\psi_{i}\right|$ is spectral decomposition of $\rho, \lambda_{i} \geq 0$ and $\sum_{i=1} \lambda_{i}=1$. When $\rho$ is pure, Eq. (\ref{F}) reduces to the variance of $H$, specifically $\mathcal{F}=$ $4 {\rm Var}_{|\alpha, m, n\rangle}(\hat{H})$. Here
 \begin{equation}
{\rm Var}_{|\alpha, m, n\rangle}(\hat{H})=
\left\langle\alpha, m, n\left|H^{2}\right| \alpha, m, n\right\rangle-4|\langle\alpha, m, n|H| \alpha, m, n\rangle|^{2}.
 \end{equation}
Let us now consider a bipartite quantum state $\rho_{A B}$ in the Hilbert space ${\cal H} ={\cal H}_{A} \otimes{\cal H}_{B}$. We
assume that the dynamics of the first subsystem is subjected to the local phase shift transformation $e^{-i \theta H_{A}}$, with $H_{A}=H_{a} \otimes \openone_{B}$ the local Hamiltonian. Therefore, QFI reduces to local quantum Fisher information (LQFI) as
\begin{equation}
\mathcal{F}\left(\rho, H_{A}\right)=\operatorname{Tr}\left(\rho H_{A}^{2}\right)-\sum_{i \neq j} \frac{2 \lambda_{i}\lambda_{j}}{\lambda_{i}+\lambda_{j}}\left|\left\langle\psi_{i}\left|H_{A}\right| \psi_{j}\right\rangle\right|^{2}.\label{21}
\end{equation}
Local quantum Fisher information is now recongnized as a discord like quantifier of quantum correlation in a bipartite quantum system \cite{Kim2018}. The local quantum Fisher information $Q\left(\rho\right)$ is defined as the minimum quantum Fisher information over all local Hamiltonians $H_{A}$ acting on the subsystem $A$ \cite{SlaouiB2019}, i.e.,
\begin{equation}
Q(\rho)=\min _{H_{A}} F\left(\rho, H_{A}\right).
\end{equation}
Any local Hamiltonian can be written $H_{a}=\vec{\sigma} \cdot \vec{r}$, with $|\vec{r}|=1$ and $\vec{\sigma}=\left(\sigma_{x}, \sigma_{y}, \sigma_{z}\right)$ are the Pauli matrices and one
has $\operatorname{Tr}\left(\rho H_{A}^{2}\right)=1$ so that the second term in (\ref{21}) can be written as
\begin{align}
&\sum_{i \neq j} \frac{2 \lambda_{i} \lambda_{j}}{\lambda_{i}+\lambda_{j}}|\left\langle  \psi_{i}\right| H_{A}\left|\psi_{j}\right\rangle|^{2}\notag\\&=\sum_{i \neq j} \sum_{l, k=1}^{3} \frac{2 \lambda_{i} \lambda_{j}}{\lambda_{i}+\lambda_{j}}\left\langle\psi_{i}\right|\sigma_{l} \otimes \openone_{B}\left| \psi_{j}\right\rangle \left\langle \psi_{j}\right| \sigma_{k} \otimes \openone_{B}\left| \psi_{i}\right\rangle \notag\\&
=\vec{r}^{\dagger} \cdot M \cdot \vec{r},
\end{align}
where the elements of the $3 \times 3$ symmetric matrix $M$ are given by
\begin{equation}
M_{l k}=\sum_{i \neq j} \frac{2\lambda_{i} \lambda_{j}}{\lambda_{i}+\lambda_{j}}\left\langle\psi_{i}\left|\sigma_{l} \otimes \openone_{B}\right| \psi_{j}\right\rangle\left\langle\psi_{j}\left|\sigma_{k} \otimes \openone_{B}\right| \psi_{i}\right\rangle.
\end{equation}
The minimal value of $F\left(\rho, H_{A}\right)$ is obtained for maximal value of the quantity $\vec{r}^{\dagger} \cdot M \cdot \vec{r}$ over all unit
vectors $\vec{r}$. The maximum value coincides with the maximal eigenvalue of the matrix $M$. Hence, the
minimal value of local quantum Fisher information $Q\left(\rho\right)$ is
\begin{equation}
Q\left(\rho\right)=1-\lambda_{\max}\left(M\right),\label{LQFI}
\end{equation}
where $\lambda_{\max }$ denotes the maximal eigenvalue of the symmetric matrix $M$. The density operator $\rho_{12}$ (\ref{Matrix12}), derived in the second partitioning scheme, is a two-qubit state of rank two. The non-vanishing eigenvalues read as
\begin{equation}
\begin{array}{l}
	\lambda_{3}=- N^{2}\left(p^{2}-p^{n-2} \cos (m \pi)\right)\left(-1+p^{2}\right)/p^{2}, \\
	\lambda_{4}= N^{2}\left(p^{2}+p^{n-2} \cos (m \pi)\right)\left(1+p^{2}\right)/p^{2}.
\end{array}
\end{equation}
The eigenstates associated with zero eigenvalues are
\begin{equation}
\left|\psi_{1}\right\rangle=\sqrt{\frac{(1+p)^{2}}{2\left(1+p^{2}\right)}}\left(1,0,0, \frac{1-p}{1+p}\right),
\end{equation}
\begin{equation}
\left|\psi_{2}\right\rangle=\frac{\sqrt{2}}{2}(0,1,-1,0),
\end{equation}
and those associated with $\lambda_{3}$ and $\lambda_{4}$ are respectively given by
\begin{equation}
\left|\psi_{3}\right\rangle=\frac{\sqrt{2}}{2}(0,1,1,0),
\end{equation}
\begin{equation}
\left|\psi_{4}\right\rangle=-\sqrt{\frac{(1-p)^{2}}{2\left(1+p^{2}\right)}}\left(1,0,0, \frac{1+p}{1-p}\right).
\end{equation}
By replacing the expressions of the eigenvectors and eigenvalues mentioned above in the elements of matrix $M$, we find
\begin{equation}
M_{11}=M_{22}=0,
\end{equation}
and
\begin{equation}
M_{33}=\frac{(1-p^{2})^{2}(1-(q \cos m \pi)^{2})}{2\left(1+ p^{n} \cos m \pi\right)^{2}}.
\end{equation}
Then, the analytic expression of the local quantum Fisher information is given by
\begin{equation}
Q(\rho)=\frac{2\left(1+ p^{n} \cos m \pi\right)^{2}-(1-p^{2})^{2}(1-(q \cos m \pi)^{2})}{2\left(1+ p^{n} \cos m \pi\right)^{2}}.
\end{equation}
\begin{figure}[h!]
\includegraphics[scale=0.7]{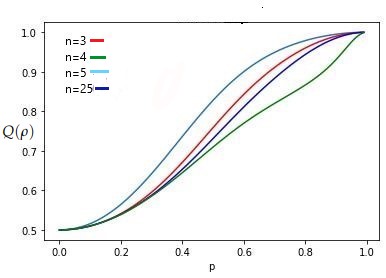}
\caption{Local quantum Fisher information as a function of the overlapping $p$ and  the number of probes $n$ with $m=0$.}\label{Fig1}
\end{figure}
\begin{figure}[h!]
	\includegraphics[scale=0.7]{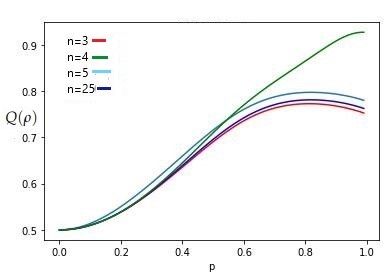}
	\caption{Local quantum Fisher information as a function of the overlapping $p$ and the number of probes $n$ with $m=1$.}\label{Fig2}
\end{figure}

The  local quantum Fisher information on Fig.\ref{Fig1} increases as $p$ increases for different values of $n$. The  maximal value of local quantum fisher information is reached for $p \longrightarrow 1$  in symmetric state where $m=0$. In this case, the corresponding states hold the maximum amount of quantum correlations and the best parameter estimation is achieved. This shows the role of quantum correlations in quantum metrology.\par

In Fig.\ref{Fig2}, the  maximal value of local quantum Fisher information is reached for $p\rightarrow1$ with $m= 1$ antisymmetric state. This limit corresponds to $W_{n}$ type states and to get estimation efficiency, it's important to use the state with $n = 4$. The case $m=0$ and $p\rightarrow1$ provides the best estimation efficiency.

\section{Local quantum uncertainty for bipartite coherent states}\label{III}
Local quantum uncertainty (LQU) is a measure of quantum correlations which captures purely quantum part in a given quantum state by applying local measurements on one part of quantum state. This measure has been defined recently for $2 \otimes d$ quantum systems \cite{Girolami2013}. Moreover, the analytical calculation method for bipartite and tripartite quantum systems are found in references \cite{Slaoui2018} and \cite{Slaoui20191}, respectively. In fact, LQU is a quantum discord-type measure and for certain quantum states, quantum discord (QD) and LQU give the same amount of quantum correlation, whereas for some other states they are different. The benefit of  LQU over quantum discord is  that to compute LQU we only need to find the maximum eigenvalue of a symmetric $3\times3$ matrix. This task is easy enough in comparison to a complex minimization process on parameters linked to the measurements. LQU is defined as the minimum skew information which is obtained via local measurement on qubit part only, that is,
\begin{equation}
\mathcal{U}(\rho) \equiv \min _{K_{A}} \mathcal{I}\left(\rho, K_{A} \otimes \openone_{B}\right),\label{LQU}
\end{equation}
where $K_{A}$ is a hermitian operator (local observable) on subsystem $A$, and $\mathcal{I}$ is the skew information of the density operator $\rho$ which is defined as \cite{Luo2003,Wigner1963}
\begin{equation}
\mathcal{I}\left(\rho, K_{A} \otimes \openone_{B}\right)=-\frac{1}{2} \operatorname{Tr}\left(\left[\sqrt{\rho}, \mathrm{K}_{\mathrm{A}} \otimes \openone_{\mathrm{B}}\right]^{2}\right).
\end{equation}
The skew information provides an analytical tool to quantify the information content in the state $\rho$ with respect to the observable $K_{a}$, the symbol $[\cdot, \cdot]$ stands for the commutator operator. The information content of $\rho$ about $K_{a}$ is here quantified by how much the measurement of $H_{a}$ on the state is uncertain. Another interesting quantum correlation quantifier in terms of skew information has been proposed in ref.\cite{Luo2012}, in which the global information content via local observables between state $\rho_{AB}$ and state $\rho_{A}\otimes\rho_{B}$ can be interpreted as a quantum correlation amount of the bipartite quantum system $\rho_{AB}$. Furthermore, the skew information is bounded by the quantum Fisher information in the phase estimation protocol and the quantum Cramer-Rao bound could be written in terms of the skew information \cite{Luo2003}. On the other hand, if it is a mixture of eigenvectors of $K_{A}$, the uncertainty is only due to imperfect knowledge of the state and the measurement outcome is certain, only if the state is an eigenvector of $K_{A}$. The closed form of the LQU for $2\times 2$ quantum systems is
\begin{equation}
\mathcal{U}(\rho)=1-\max \left\{\omega_{11},\omega_{22}, \omega_{33}\right\},
\end{equation}
where $\omega_{11},\omega_{22}$ and $\omega_{33}$ are the eigenvalues of the $3 \times 3$ matrix $W$ whose matrix elements are defined by
\begin{equation}
\omega_{i j} \equiv \operatorname{Tr}\left\{\sqrt{\rho}\left(\sigma_{i} \otimes \openone_{B}\right) \sqrt{\rho}\left(\sigma_{j} \otimes \openone_{B}\right)\right\},\label{W}
\end{equation}
with $i, j=1,2,3$ and the $\sigma_{i}$'s represent the Pauli matrices. Indeed, the computation of the elements of the matrix $W$ (\ref{W}) is too long. For the so-called X states, for which the density matrix contains the non-zero entries only along the diagonal and anti-diagonal, the analytical expression of local quantum uncertainty was derived in Ref. \cite{Slaoui2018}. By using this method and after some simplifications, the explicit expressions of the non-vanishing elements of the matrix $W$ are given by
\begin{equation}
\omega_{11} = \frac{N^{2}(1-p^{2})^{2}(1-(q \cos m \pi)^{2})}{\sqrt{2\left(1- q^{2} \cos m \pi\right)^{2}(1-p^{4})}},
\end{equation}
\begin{equation}
\omega_{22} = \frac{N^{2}(1-p^{2})^{2}(1-(q \cos m \pi)^{2})}{\sqrt{2\left(1- q^{2} \cos m \pi\right)^{2}(1-p^{4})}}p^{2},
\end{equation}
and
\begin{equation}
\omega_{33} = \frac{1}{2} - \frac{N^{2}( 1 + p^{n+2}\cos m \pi - 3(p^2 + p^{n} \cos m \pi))}{1+p^{2}}.
\end{equation}
It is obvious that $\omega_{11}\geq\omega_{22}$ (since $\omega_{11}=p^{2}\omega_{22}$), then the local quantum uncertainty is
\begin{equation}
\mathcal{U}(\rho)=1-\max \left\{\omega_{11}, \omega_{33}\right\}.
\end{equation}
\begin{figure}[h!]
	\includegraphics[scale=0.95]{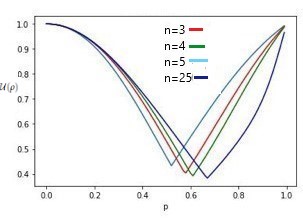}
	\caption{The values of Local quantum uncerntaity as a function of the overlap $p$  and the number of probes $n$ for symmetric state.}\label{Fig3}
\end{figure}  
\begin{figure}[h!]
	\includegraphics[scale=0.7]{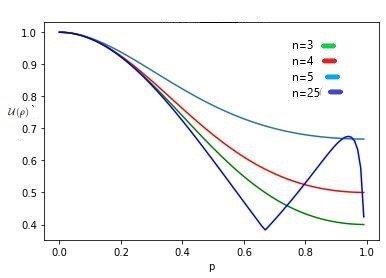}
	\caption{Local quantum uncerntaity as a function of the overlap $p$  and  the number of probes n  for antisymmetric state.}\label{Fig4}
\end{figure}

The behavior of LQU versus the overlapping $p$ and the number of probes $n$ for symmetric state and antisymmetric state is plotted in Fig.(\ref{Fig3}) and Fig.(\ref{Fig4}). As it can be seen from these figures, the  maximal value of local quantum uncertainty  is reached for $p\rightarrow1$. This limit corresponds to the ground state  and for $p\rightarrow0$, the state reduces to $GHZ_{n}$ type states.\par

For antisymmetric states (see Fig.\ref{Fig4}), the maximum value of the local quantum uncertainty is reached for $p\rightarrow0$. This limit corresponds to $GHZ_{n}$ type states regardless the number of probes. 
\section{Dynamics of LQFI and LQU under dephasing channel}\label{V}
Decoherence arises as a consequence of the interaction between the system, the apparatus and the environment \cite{Paz1999,Joos1985}. In this section, we study the behavior of LQFI and LQU under the effect of decoherence phenomena. To describe these decoherence effects, we denote by $\mathcal{E}$ the quantum operation that maps the physical state given by the density operator $\rho$ to another state $\rho^{\prime}$ which depends on decoherence probability $\gamma$ such that
\begin{equation}
\rho^{\prime}=\mathcal{E}(\rho)=\sum_{m} K_{m} \rho K_{m}^{\dagger},
\end{equation}
where the $K_{m}$ are Kraus operators satisfying the completeness relation $\sum_{m} K_{m}^{\dagger}K_{m}=\openone$. A quantum operation describing the environment noise is sometimes called a quantum channel \cite{Lloyd1997,Devetak2005}. A good practical example of a quantum channel is the dephasing channel. This case is especially informative, since it provides a revealing illustration of decoherence in physically realistic situations. An individual isometric representation of this channel is
\begin{align}
	|0\rangle_{A} \mapsto \sqrt{1-\gamma}|0\rangle_{A}
	\otimes|0\rangle_{E}+\sqrt{\gamma}|0\rangle_{A} \otimes|1\rangle_{E}, \\
|1\rangle_{A} \mapsto \sqrt{1-\gamma}|1\rangle_{A}
	\otimes|0\rangle_{E}+\sqrt{\gamma}|1\rangle_{A} \otimes|2\rangle_{E}.
\end{align}
Under this effect, the qubit $A$ does not make any transitions in the reference basis $\left\lbrace\left|0\right\rangle_{A},\left|1\right\rangle_{A} \right\rbrace$ (this basis is the only one in which bit flips never occur). But the environment is occasionally scattered out of the qubit with probability $\gamma$; evolving into state $\left|1\right\rangle_{E}$ if the qubit $A$ is in state $\left|0\right\rangle_{A}$ and into state $\left|2\right\rangle_{E}$ if the qubit $A$ is in state $\left|1\right\rangle_{A}$. Evaluating the partial trace on the basis of the environment $\left\{|0\rangle_{E},|1\rangle_{E},|2\rangle_{E}\right\}$, we get the following Kraus operators
\begin{equation}
\boldsymbol{K}_{1}=\sqrt{\gamma}\left(\begin{array}{ll}
	1 & 0 \\
	0 & 0
\end{array}\right)=\frac{\sqrt{\gamma}}{2}\left(\boldsymbol{I}+\boldsymbol
{\sigma}_{3}\right),
\end{equation}
and
\begin{equation}
\boldsymbol{K}_{2}=\sqrt{\gamma}\left(\begin{array}{ll}
	0 & 0 \\
	0 & 1
\end{array}\right)=\frac{\sqrt{\gamma}}{2}\left(\boldsymbol{I}-
\boldsymbol{\sigma}_{3}\right).
\end{equation}
Therefore, the dephasing channel maps the state $\rho_{12}$ (\ref{Matrix12}) to
\begin{equation}
	\mathcal{E}(\boldsymbol{\rho_{12}})\equiv\rho_{12}^{DC}=\left(1-\frac{1}{2} \gamma\right) \boldsymbol{\rho_{12}}+\frac{1}{2} \gamma
	\boldsymbol{\sigma}_{3} \rho_{12} \boldsymbol{\sigma}_{3},
\end{equation}
where $\gamma=1-e^{-\Gamma t}$ and $\Gamma$ indicates the decay rate. Alternatively, this channel can be described by saying that $\sigma_{3}$ is applied with probability $\gamma / 2$ and nothing happens with probability $(1-\gamma / 2)$. The density matrix evolves as
{\small \begin{equation}
	\rho_{12}^{DC}\left(t\right)=N^{2}\left(\begin{array}{cccc}
		\eta_{a} & 0 & 0 & (1-\gamma)\eta_{+} \\
		0 & (1-\gamma)\eta_{-} & (1-\gamma)\eta_{-} & 0 \\
		0 & (1-\gamma)\eta_{-} & (1-\gamma)\eta_{-} & 0 \\
		(1-\gamma)\eta_{+} & 0 & 0 & \eta_{b}
	\end{array}\right).\label{rhoDC}
\end{equation}}
The corresponding eigenvalues are read as
\begin{align}
\lambda_{1,4}^{DC}=&\frac{N^{2}}{2}\left(1+q \cos\left(m \pi\right)\right)\left(1+p^{2}\right)\notag\\&\times\left( 1+\sqrt{1 \pm\frac{(2\gamma - \gamma^{2})(1-p^{2})^{2}}{(1+p^2)^{2}}}\right),
\end{align}
\begin{equation}
\lambda_{2,3}^{DC}= \frac{N^{2}}{2}\left(1-q \cos\left(m \pi\right) \right)\left(1-p^{2}\right)\left(1 \mp\left(\gamma - 1\right)\right).
\end{equation}
The corresponding eigenvectors are read as
\begin{equation*}
\left|\psi_{1}^{DC}\right\rangle=\left(\sqrt{\frac{-\chi_{+}^2}{1+\chi_{+}^2}},0,0,\sqrt{\frac{1}{1+\chi_{+}^2}}\right)^{T},
\end{equation*}
\begin{equation*}
\left|\psi_{2}^{DC}\right\rangle=\left(\sqrt{\frac{-\chi_{-}^2}{1+\chi_{-}^2}},0,0,\sqrt{\frac{1}{1+\chi_{-}^2}}\right)^{T},
\end{equation*}
\begin{equation*}
\left|\psi_{3}^{DC}\right\rangle=\left(\sqrt{\frac{1}{1+\delta^2}},0,0,-\sqrt{\frac{\delta^2}{1+\delta^2}}\right)^{T},
\end{equation*}
\begin{equation*}
\left|\psi_{4}^{DC}\right\rangle=\left(\sqrt{\frac{1}{1+\beta^2}},0,0,-\sqrt{\frac{\beta^2}{1+\beta^2}}\right)^{T},
\end{equation*}
where
\begin{equation}
	\chi_{\pm}= N^2\frac{(1-p^2)^2(2+\gamma^2-2\gamma)-\xi_{\pm}(1+p^2)\left(4p+\xi_{\pm}\left(1+p^{2}\right) \right)}{2(1-\gamma)(1-p^2)\left((1+p)^2 + (1+p^2)\xi_{\pm}\right)},
\end{equation}
and
\begin{equation}
\beta= \frac{1- N^2\gamma+q\cos m \pi\left(1+ N^2\gamma\right)}{(1+q\cos m \pi)(1-p)},
\end{equation}
\begin{equation}
	\delta=1-\frac{p}{q\cos m\pi\left(2+p-\gamma\left(1+p \right)\right) }.
\end{equation}
\subsection{LQFI under dephasing channel}
Here, we study the dynamics of LQFI under dephasing channel for multipartite coherent states (\ref{rhoDC}). From the equation (\ref{LQFI}), the expression of LQFI writes 
\begin{equation}
	Q\left(\rho_{12}^{DC}\right)=1-\lambda_{\max}^{DC}\left(M^{DC}\right).
\end{equation}
In order to compute the $M^{DC}$ matrix, we use the eigenvalues and eigenvectors of state $\rho^{DC}$ (\ref{rhoDC}). After some simplifications, all the elements of the matrix $M^{DC}$ are equal to zero except the element $M_{33}^{DC}$ which takes the form
\begin{widetext}
\begin{align}
&M_{33}^{DC}= \frac{2(1-p^4)\xi_{+}\gamma N^{2}(1-(q\cos m \pi)^2)(\beta-\chi_{+})^2}{(1+\beta^2)(1+\chi_{+}^2)(\gamma (1-q\cos m \pi)(1-p^2) + (1+p^2)(1+q\cos m\pi)\xi_{+})}+\frac{\gamma\left(1-p^2\right) \left(1-q\cos m\pi\right)\left(\beta^2 -1\right)}{2\left(\beta^2+1 \right)}\notag\\&+  \frac{2\gamma N^{2}(1-p^4)\xi_{-}(1-(q\cos m \pi)^2)(\beta-\chi_{-})^2}{(1+\chi_{-}^2)(1+\beta^2)(\gamma (1-q\cos m \pi)(1-p^2) + (1+p^2)(1+q\cos m\pi)\xi_{-})}+ \frac{N^{2}(1+q\cos m\pi)\gamma(2-\gamma)(1-p^2)^2 (\chi_{+}\chi_{-}-1)^2}{2(1+p^2) (1+\beta^2)(1+\chi_{-}^2)},
\end{align}
with

\begin{equation}
	\xi_{\pm} = \left(1\pm\sqrt{1 - \frac{\gamma(2 - \gamma)(1-p^2)^2}{(1+p^2)^2}}\right).
\end{equation}	
	\begin{figure}[h!]
		{{\begin{minipage}[b]{.25\linewidth}
					\centering
					\includegraphics[scale=0.48]{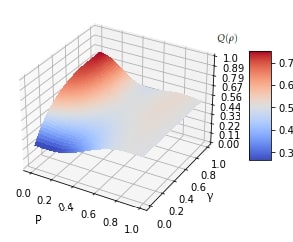}\vfill
					$n=3$
				\end{minipage}\hfill
				\begin{minipage}[b]{.25\linewidth}
					\centering
					\includegraphics[scale=0.48]{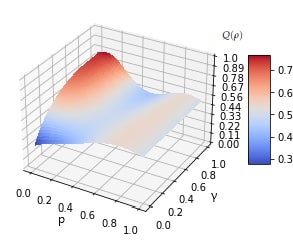}\vfill
					$n=4$
				\end{minipage}\hfill
			\begin{minipage}[b]{.25\linewidth}
				\centering
				\includegraphics[scale=0.48]{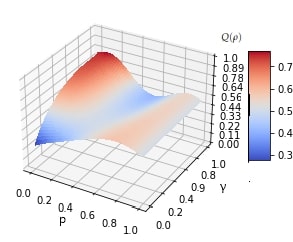}\vfill
				$n=5$
			\end{minipage}\hfill
				\begin{minipage}[b]{.25\linewidth}
					\centering
					\includegraphics[scale=0.48]{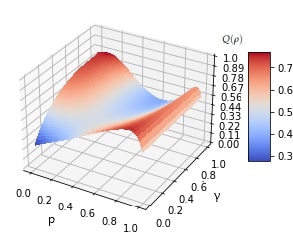}\vfill
					$n=25$
		\end{minipage}}}
		\caption{Local quantum Fisher information for symmetric state $m=0$ under dephasing channel effect.}\label{Fig5}
	\end{figure}
	\begin{figure}[h!]
		{{\begin{minipage}[b]{.25\linewidth}
					\centering
					\includegraphics[scale=0.48]{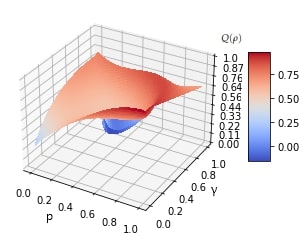}\vfill
					$n=3$
				\end{minipage}\hfill
				\begin{minipage}[b]{.25\linewidth}
					\centering
					\includegraphics[scale=0.48]{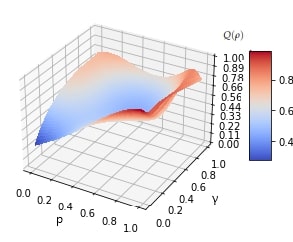}\vfill
				$n=4$
				\end{minipage}\hfill
				\begin{minipage}[b]{.25\linewidth}
					\centering
					\includegraphics[scale=0.48]{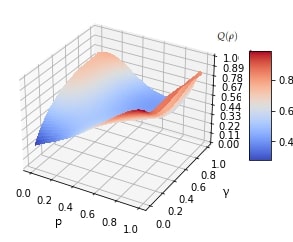}\vfill
					$n=5$
				\end{minipage}\hfill
				\begin{minipage}[b]{.25\linewidth}
					\centering
					\includegraphics[scale=0.48]{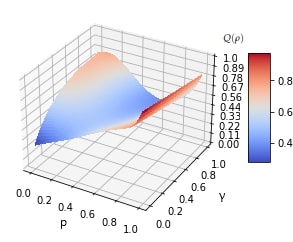}\vfill
					$n=25$
		\end{minipage}}}
		\caption{Local quantum Fisher information for antisymmetric state $m=1$ under dephasing channel effect.}\label{Fig6}
	\end{figure}
\end{widetext}
In order to observe the influence of the decoherence probability $\gamma$ and the overlapping $p$ that specify the coherent states on the local quantum Fisher information for mixed states, we plotted numerically in Fig.(\ref{Fig5}) the variation of LQFI present in the symmetric multipartite coherent states ($m$ even) with respect to the overlapping $p$ and the phase shift probability $\gamma$ for fixed values of the probe numbers $n$. The main result obtained from this figure is that the amount of LQFI is strongly dependent on the value of $\gamma$. It reaches the maximal value when the mixed state is GHZ-type and totally dephased (i.e. $\gamma\equiv1$). Moreover, LQFI shows a robust behavior of quantum correlations against decoherence when the coherent state is ground state, but the GHZ-type coherent state still remains the best for quantum metrology estimation protocols. As an important remark, increasing the probe numbers $n$ plays an important role in the development of the LQFI robustness as it can be seen from Fig.(\ref{Fig5}) when $n=25$, meaning that it tends to a constant value for different values of $\gamma$ when $p=1$, but still does not represent the maximal value of LQFI.\par 

The Plots in Fig.(\ref{Fig6}) illustrate the behavior of local quantum Fisher information versus the overlapping $p$ and the dephasing probability $\gamma$ for antisymmetric multipartite coherent states ($m$ odd). Initially, the LQFI declines to a minimum and increases thereafter to maintain a very slow growth. As results, we conclude that the GHZ-type state does not perform well on the anti-symmetric states and does not reach the values of the first case ($m$ even). Indeed, the values of LQFI increase, for all values of $\gamma$, when the state is of Werner type, i.e. $p\rightarrow1$. Furthermore, one can assert also that the increase of the probes numbers plays an important role regarding the robustness of the quantum correlations, especially when $n= 25$.\par
In addition to the behavior and related interpretation given above, we show that the correlation, which is quantized by LQFI, is more robust against the dephasing channel in the antisymmetric state compared to the correlation in the symmetric state.

\subsection{LQU under dephasing channel}
Using the expression (\ref{LQU}), we calculate the matrix $W^{DC}$  to compute the local quantum uncertainty in presence of  decoherence effects. We obtain $W^{DC}\equiv{\rm diag}\left\lbrace w_{11}^{DC},w_{22}^{DC},w_{33}^{DC}\right\rbrace$,
where the expressions of $w_{11}^{DC}$, $w_{22}^{DC}$ and $w_{33}^{DC}$ are given by
\begin{widetext}
 \begin{align}
	w_{11}^{DC}=& \frac{2N^{2}(1-p^{4})(1-(q\cos m\pi)^{2})(1+\sqrt{\gamma(2-\gamma)})\Delta}{\sqrt{(1-p^{4})(1-(q\cos m\pi)^{2}}(\sqrt{\gamma}+\sqrt{2 - \gamma})(\sqrt{1 + \Lambda}+\sqrt{1-\Lambda})}\notag\\&+\frac{ N^{2}(1-\gamma)^{2}((1-p^{2})^{2}-16(q\cos m\pi)^{2})}{\sqrt{(1-p^{4})(1-(q\cos m\pi)^{2}}(\sqrt{\gamma}+\sqrt{2 - \gamma})(\sqrt{1 + \Lambda}+\sqrt{1-\Lambda})},
\end{align}
\begin{align}
		w_{22}^{DC}=&\frac{2N^{2}(1-p^{4})(1-(q\cos m\pi)^{2})(1+\sqrt{\gamma(2-\gamma)})\Delta}{\sqrt{(1-p^{4})(1-(q\cos m\pi)^{2}}(\sqrt{\gamma}+\sqrt{2 - \gamma})(\sqrt{1 + \Lambda}+\sqrt{1-\Lambda})}\notag\\&-\frac{ N^{2}(1-\gamma)^{2}((1-p^{2})^{2}-16(q\cos m\pi)^{2})}{\sqrt{(1-p^{4})(1-(q\cos m\pi)^{2}}(\sqrt{\gamma}+\sqrt{2 - \gamma})(\sqrt{1 + \Lambda}+\sqrt{1-\Lambda})},
\end{align}
and
\begin{align}
w_{33}^{DC}=&\frac{2N^{2}(1+q\cos m\pi)^{2}p^{2}-(1-\gamma)^{2}((1-p^{2})+4 q\cos m\pi)^{2}}{(1+q\cos m\pi)(1+p^{2})\Delta}+\frac{N^{2}}{2}4(1+q\cos m\pi)(1+p^{2})\Delta\notag\\&+\frac{N^{2}}{2}(1-q\cos m\pi)(1-p^{2})(1 + \sqrt{2\gamma - \gamma^{2}})-\frac{N^{2}(1-\gamma)^{2}((1-p^{2})-4q\cos m\pi)^{2}}{2(1+q\cos m\pi)(1-p^{2})\Delta},
\end{align}
with
\begin{equation}
	\Lambda=\frac{\sqrt{\left(1+p^{2}\right)^{2}-\left(1-p^{2}\right)^{2}\left(2-\gamma\right)^{2}\gamma^{2}} }{1+p^{2}},\hspace{1cm}{\rm and}\hspace{1cm}\Delta=1+\frac{\left(1-p^{2}\right)\sqrt{\gamma\left(2-\gamma\right)} }{1+p^{2}}.
\end{equation}
Depending on the largest value between $w_{11}^{DC}$ and $w_{33}^{DC}$, 
the LQU writes
\begin{equation}
\mathcal{U}\left(\rho\right)^{DC}=1-\max\left\{w_{11}^{DC}, w_{33}^{DC}\right\}.
\end{equation}
\begin{figure}[h!]
		{{\begin{minipage}[b]{.25\linewidth}
					\centering
					\includegraphics[scale=0.48]{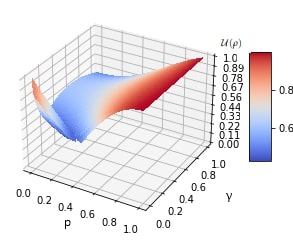}\vfill
					$n=3 $
				\end{minipage}\hfill
				\begin{minipage}[b]{.25\linewidth}
					\centering
					\includegraphics[scale=0.48]{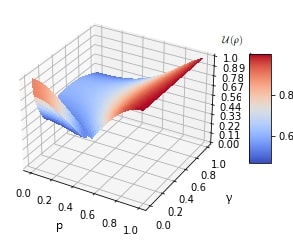}\vfill
					$n=4 $
				\end{minipage}\hfill
				\begin{minipage}[b]{.25\linewidth}
					\centering
					\includegraphics[scale=0.48]{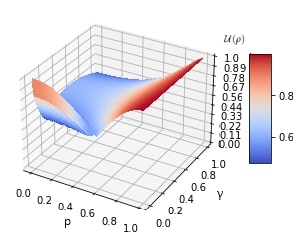}\vfill
					$n=5 $
				\end{minipage}\hfill
				\begin{minipage}[b]{.25\linewidth}
					\centering
					\includegraphics[scale=0.48]{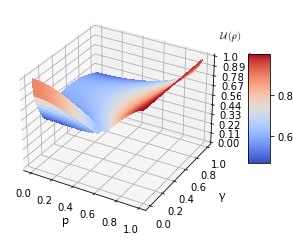}\vfill
					$n=25$
		\end{minipage}}}
		\caption{Local quantum uncertainty for symmetric state $m=0$ under dephasing channel effect.}\label{Fig7}
	\end{figure}

	\begin{figure}[h!]
	{{\begin{minipage}[b]{.25\linewidth}
				\centering
				\includegraphics[scale=0.48]{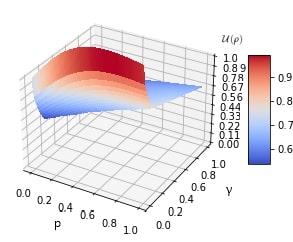}\vfill
				$n=3$
			\end{minipage}\hfill
			\begin{minipage}[b]{.25\linewidth}
				\centering
				\includegraphics[scale=0.48]{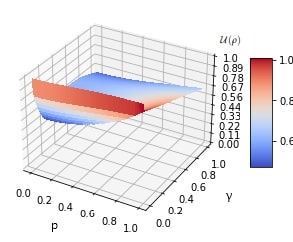}\vfill
				$n=4$
			\end{minipage}\hfill
			\begin{minipage}[b]{.25\linewidth}
				\centering
				\includegraphics[scale=0.48]{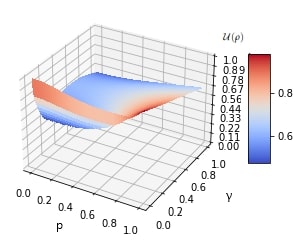}\vfill
				$n=5$
			\end{minipage}\hfill
			\begin{minipage}[b]{.25\linewidth}
				\centering
				\includegraphics[scale=0.48]{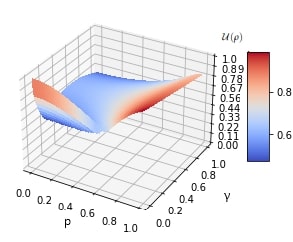}\vfill
				$n=25$
	\end{minipage}}}
	\caption{Local quantum uncertainty for antisymmetric state $m=1$ under dephasing channel effect.}\label{Fig8}
\end{figure}
\end{widetext}

Here, we analyze the sensitivity of the overlapping $p$, the number of probes $n$ and the phase shift probability $\gamma$ on the behavior of the local quantum uncertainty, and the results are reported in figure \ref{Fig7} for symmetric multipartite coherent states and in figure \ref{Fig8} for antisymmetric multipartite coherent states.\par

By controlling the number of probes, it is clear that a large number of photons induces a decrease of the local quantum uncertainty quantities, where interestingly the best precision in quantum estimation is given at small values of $n$. It can be seen that the LQU decreases with increasing values of the phase shift probability. Hence, the phase shift modifies the initial correlations of the multipartite Glauber coherent states and the information may be lost if a dephasing channel is present. Furthermore, when we compare these findings with the results reported in figures \ref{Fig5} and \ref{Fig6}, the destructive effects of the phase shift are much stronger for LQU than those for quantum correlations measured by LQFI. This confirms the robustness of LQU in comparison with LQFI against the dephasing channel effect. Concerning the symmetric coherent state types which gives us a better precision of the estimated parameters, we notice that the maximal value of LQU is obtained when the multipartite superposition approaches to the ground state. More interestingly, the LQU values decrease when the mixed state is GHZ-type (see Fig.(\ref{Fig7})). For the case of antisymmetric coherent states, we observe that both GHZ and Werner states are fragile against the decoherence effect, and we see that there's a slight increase in robustness when we increase the number of probes (see Fig.(\ref{Fig8})).\par

On the other hand, as displayed both in Fig.(\ref{Fig7}) for the symmetric state and in Fig.(\ref{Fig8}) for the antisymmetric state, when we increase the number of probes $n$, the local quantum uncertainty decreases. The local quantum uncertainty is therefore maximized in the case where $n=3$.

\section{Concluding Remarks}\label{IV}
To close this paper, we recall that the main motivation of this paper is to examine the role of pairwise quantum correlations contained in multipartite Glauber coherent states, captured by local quantum Fisher information and local quantum uncertainty, in improving the precision in quantum metrology protocols, especially in estimation protocols for an unknown phase shift. We divided the complete system (containing $n$ particles) into two parts in which one contains $k$ particles and the second consists of the remaining $n-k$ particles resulting in a pure state. Another bi-partitioning scheme consists of tracing $n-2$ particles and gives a mixed state system.\par

In this way, we explicitly derived the analytical expressions of local quantum Fisher information and local quantum uncertainty in a multipartite system including coherent states by an appropriate qubit mapping for all possible bi-partitions. Our comparative study shows that the local quantum Fisher information reaches the best value of the estimation protocols for the symmetric case (i.e., $m=0$) and the coherent state is of the ground state type, whereas for the antisymmetric states, the local quantum Fisher information increases when the number of probes increases and when the coherent state is a Werner state type. Moreover, the local quantum uncertainty reaches the best value for the estimation protocols when the coherent state can be viewed as Greenberger-Horne-Zeilinger for both symmetric and antisymmetric states regardless of the number of probes.\par

The decoherence effects on the dynamics of the quantum correlations are fully investigated for two kinds of multipartite Glauber coherent states. Our study proves that the behavior of quantum correlations in not influenced only by the phase shift probabilit $\gamma$, but also by the number of probes $n$ where the quantum correlations decreases with the increase of the number of probes. Additionally, under a noisy channel, LQU has the best robustness against the phase shift channel, especially when the multipartite coherent states behave like a multipartite ground state. Whenever the state is symmetric, LQU also retains its power. Therefore, it one concludes that, for improving the accuracy of quantum metrology protocols, especially when there is interaction with the environment, LQU is the best quantifier of quantum correlations. Besides, we cannot ignore also the performance of LQFI under the effect of decoherence when $p\rightarrow0$ on the GHZ-type state. We also conclude that the ground state of multipartite coherent states is also a good resource for quantum protocols like entanglement states and there is a small effect on the number of probes. This effect depends on the state type and the quantum correlations quantifier. To give an example, LQFI is better to increase the number of probes but for LQU, it is clear to use it only when $n=3$ or $n=4$.\par
This study regarding the role of quantum correlations beyond entanglement in quantum metrology may stimulate experimental studies focused on the realization of high precision protocols in which the information is encoded in coherent states.

\end{document}